\documentclass[aps,pra,superscriptaddress,showpacs,twocolumn,10pt,longbibliography,floatfix]{revtex4-1}
\usepackage{graphicx}
\usepackage{bm,amsmath,amsfonts}
\usepackage{amssymb}
\usepackage{amsbsy}
\usepackage{verbatim}
\usepackage{color}
\usepackage{hyperref}
\usepackage[american]{babel}
\usepackage[utf8]{inputenc}
\usepackage[T1]{fontenc}
\usepackage{lmodern}

\begin{hyphenrules}{american}
\end{hyphenrules}

\begin{document}

\author{I. Hans}
\email{ihans@physnet.uni-hamburg.de}
\affiliation{Zentrum f\"ur Optische Quantentechnologien,
  Universit\"at Hamburg, Luruper Chaussee 149, 22761 Hamburg, Germany}

\author{J. Stockhofe}
\email{jstockho@physnet.uni-hamburg.de}
\affiliation{Zentrum f\"ur Optische Quantentechnologien,
  Universit\"at Hamburg, Luruper Chaussee 149, 22761 Hamburg, Germany}

\author{P. Schmelcher}
\email{pschmelc@physnet.uni-hamburg.de}
\affiliation{Zentrum f\"ur Optische Quantentechnologien,
  Universit\"at Hamburg, Luruper Chaussee 149, 22761 Hamburg, Germany}
\affiliation{The Hamburg Centre for Ultrafast Imaging, Universit\"at Hamburg, Luruper Chaussee 149, 22761 Hamburg, Germany}

\title{Generating, dragging and releasing dark solitons \\in elongated Bose-Einstein condensates}

\begin{abstract}
We theoretically analyze quasi-one-dimensional Bose-Einstein condensates under the influence of a harmonic trap and a narrow potential defect that moves through the atomic cloud.
Performing simulations on the mean field level, we explore a robust mechanism in which a single dark soliton is nucleated and immediately pinned by the moving defect,
making it possible to drag it to a desired position and release it there. We argue on a perturbative level that a defect potential which is attractive to the atoms is suitable for holding and moving dark solitons.
The soliton generation protocol is investigated over a wide range of model parameters and its success is systematically quantified by a suitable fidelity measure,
demonstrating its robustness against parameter variations, but also the need for tight focusing of the defect potential.
Holding the soliton at a stationary defect for long times may give rise to dynamical instabilities, whose origin we explore within a Bogoliubov-de Gennes linearization analysis.
We show that iterating the generation process with multiple defects offers a perspective for initializing multiple soliton dynamics with freely chosen initial conditions. 
\end{abstract}

\maketitle
\section{Introduction}
Solitonic wave excitations that maintain their shape during propagation 
are found in a large variety of physical systems, ranging from hydrodynamics to modern telecommunication systems and even biological molecules \cite{solitons}.
The enormous technological advance of recent years has made it possible to prepare and observe solitons in ultracold atom experiments,
primarily (but not only, see e.g. \cite{Yefsah2013}) in condensed bosonic ensembles near zero temperature \cite{Pethick_book,book_kevrekidis}.
On the mean-field level, a Bose-Einstein condensate (BEC) is described by the Gross-Pitaevskii equation (GPE) \cite{Pethick_book}, a nonlinear Schrödinger equation with cubic nonlinearity
induced by the interatomic interaction. In one spatial dimension (1D), this equation is well-known to feature dark (bright) solitons for defocusing (focusing) nonlinearities, respectively \cite{book_kevrekidis,DJF}.
Experimentally, a highly elongated quasi-1D regime can be reached by tightly confining the atoms in the radial direction, effectively freezing out the transverse dynamics.
Early experiments succeeded in preparing bright \cite{L.Khaykovich05172002,Strecker2002} and dark \cite{PhysRevLett.83.5198,Science.287.5450} matter-wave solitons.
Dark solitons (which are in the focus of this work) are characterized by a localized density minimum across which the phase changes by $\pi$.
These can be created by manipulating the condensate phase \cite{PhysRevLett.83.5198,Science.287.5450, Becker2008, PhysRevLett.101.120406} or density \cite{ZacharyDutton07272001} using external potentials, see also the discussion in \cite{DJF}.
They can also form in the wake of a repulsive barrier dragged through the condensate \cite{Pavloff2002, Radouani2004,Engels2007} or in collisions of initially separated atomic clouds \cite{Weller2008}.
Still, controllably creating a dark soliton at a desired position in a Bose-Einstein condensate is a challenging task in experiments.\\
Once a soliton has formed, its dynamics can again be influenced with external potentials. 
A dark soliton in a harmonically trapped BEC performs particle-like harmonic oscillations around the trap center
\cite{PhysRevLett.84.2298,Becker2008,PhysRevLett.101.120406,Weller2008}. 
The emission of sound waves due to acceleration of a soliton has been studied in \cite{Parker2003a,Parker2003,Pelinovsky2005a,Parker2010}.
Thinking of the external force as a handle for controlling the soliton dynamics, a direct manipulation of the soliton motion with narrow potential defects has been suggested in \cite{Frantzeskakis2002},
where the interaction of a dark soliton with a pointlike impurity was analyzed within perturbation theory (see also \cite{Kivshar1994,Konotop1994,Konotop1997} and the corresponding studies for bright \cite{Herring2005} and dark-bright solitons \cite{Achilleos2011}).
Moreover, the possibility of dragging along dark solitons in a moving optical lattice potential has been demonstrated in \cite{Theocharis2005a,Theocharis2005}.
In a similar spirit, dragging of bright solitons in a discrete lattice model has been discussed recently \cite{Brazhnyi2011}, while pinning and transporting quantum vortices with focused external potentials has been shown in \cite{Tung2006,Davis2009}.\\
Here, we describe a method for controllably creating, dragging and releasing a dark soliton in a repulsively interacting Bose-Einstein condensate employing a tightly focused red-detuned laser beam (acting as an attractive potential for the atoms via the dipole force). 
Specifically, we study a trapped quasi-1D BEC under the influence of a moving Gaussian potential defect of attractive sign.
When entering the atomic cloud from outside with this defect, a dark soliton can be created and at the same time pinned by the defect, such that it can subsequently be placed and released at an arbitrary position.
Motivated by an analysis of the instantaneous energy levels in a single particle model, a similar scheme for exciting nodes in the condensate wave function by traversing it with an attractive defect has been proposed in \cite{Karkuszewski2001,Damski2001}.
A related protocol for the extraction of bright solitons from an attractive BEC has been suggested in \cite{Carpentier2008} (see also \cite{Carpentier2006}), but there the defect was not assumed to act as an external potential, 
but instead to cause a local variation of the effective atomic interaction through the mechanism of optical Feshbach resonance \cite{Theis2004}.
In our work, the defect acts as a single particle potential for the atoms. Dynamically manipulating (e.g. splitting) the entire BEC cloud with such optical ``tweezers'' is nowadays well established \cite{Gustavson2001,Boyer2006}.
If, instead, the light is focused to the comparably short length scale of a dark soliton, the same type of technology can be employed to manipulate the dynamics of a localized solitonic excitation.\\
Our presentation is structured as follows. In Sec. \ref{perturbth} we introduce the theoretical mean-field framework our study is based on and give a discussion of the results from a perturbative treatment of the potential defect, revealing in particular that a potential that is attractive to the individual atoms is effectively also attractive to a dark soliton.
In Sec.\ref{sec:generation} we show results from numerical simulations, demonstrating the robust creation, dragging and release of a dark soliton. 
We quantify the fidelity of the creation process and vary the defect parameters to explore the robustness of this protocol.
If the dark soliton is pinned to the defect for long times, a dynamical instability may occur which we address in Sec.\ref{sec:instab}, making the connection to a Bogoliubov-de Gennes linearization around the corresponding stationary solution of the GPE.
We briefly conclude and point to further perspectives in Sec.\ref{conclu} . Details of the perturbation theory are defered to appendix \ref{app}.

\section{Setup and results from perturbation theory}\label{perturbth}
We investigate a quasi-1D Bose-Einstein condensate of a single atomic species with repulsive interaction at zero temperature, described by the Gross-Pitaevskii equation \cite{Pethick_book}. 
We assume strong harmonic confinement in two spatial directions, taking the trapping potential to be of the form $V_{\text{3D}}(\mathbf r, t)=\frac{m \omega_\perp^2}{2}(y^2+z^2)+V(x,t)$,
where $m$ denotes the atomic mass, $\omega_\perp$ the frequency of the transverse oscillator potential and $V(x,t)$ models a potential in the longitudinal direction.
Assuming that the transverse dynamics is fully frozen out and thus transversally the condensate wave function remains in the oscillator ground state,
one can integrate out the $y$- and $z$-directions and is left with the effectively 1D Gross-Pitaevskii equation \cite{book_kevrekidis}
\begin{align}
-\dfrac{\hbar^2}{2m} \partial_{x}^2 \psi + V(x,t) \psi + g_{\text{1D}} |\psi|^2 \psi  = i \hbar  \partial_{t}\psi ,
\label{eq:GPEdimensions}
\end{align}
where $\psi=\psi(x,t)$ denotes the longitudinal part of the wave function and the nonlinearity coefficient $g_{\text{1D}}=2\alpha\hbar \omega_{\perp}$ with the $s$-wave scattering length $\alpha>0$ \cite{Pethick_book}.
Measuring length, time, energy and density $|\psi (x,t)|^2$ in units of $a_{\perp}=\sqrt{\hbar/m\omega_{\perp}}$, $\omega_{\perp}^{-1}$, $\hbar \omega_{\perp}$ and $(2\alpha)^{-1}$, respectively,
Eq.~(\ref{eq:GPEdimensions}) is cast into the dimensionless form
\begin{align}
-\dfrac{1}{2} \partial_{x}^2 \psi + V(x,t) \psi +  |\psi|^2 \psi = i  \partial_t \psi \label{dimlesstimedepGPE1}
\end{align}
which we will work with in the following.

The corresponding stationary equation is obtained by factorizing $\psi(x,t)=\phi(x) \exp(-i\mu t)$ with $\mu$ the chemical potential.
Our focus here will be on a longitudinal potential that consists of a static harmonic part (whose frequency $\omega_\parallel$ is small compared to that of the transverse trap)
plus a Gaussian of fixed height and width, but moving in time, i.e. in dimensionless units
\begin{align}
V(x,t)=\dfrac{1}{2} \Omega^2 x^2 + V_0 \exp \left[-\frac{1}{2} \frac{(x-x_G(t))^2}{\sigma ^2}\right] \label{potential}.
\end{align}
Here, $x_G(t)$ specifies the trajectory of the Gaussian impurity, while $\sigma$ and $V_0$ set its width and amplitude, respectively.
For convenience, the aspect ratio will be fixed to $\Omega^2=\omega_\parallel^2/\omega_\perp^2=0.04$ in the following. 
Our results can be transfered to other aspect ratios by a straightforward rescaling of Eqs.~(\ref{dimlesstimedepGPE1},\ref{potential}).
We simulate the time evolution of the condensate by propagating Eq.~(\ref{dimlesstimedepGPE1}) with a fourth-order Runge-Kutta integrator. 
The initial state is chosen to be the ground state at a given $\mu$, obtained from the stationary 1D-GPE by an adapted Newton method \cite{kelley}.
\\
The main objective of this work is the controlled formation and dragging of a dark soliton. 
By analogy with previous work on vortices in 2D \cite{Davis2009}, and the somewhat intuitive picture that a density dip such as the dark soliton may offer the possibility of pinning it with a narrow repulsive barrier in its center,
one may conjecture that $V_0 > 0$ will be the favorable parameter regime for our purposes.
It turns out, however, that this is not the case and an attractive Gaussian impurity with $V_0 < 0$ is much more adequate to drag along the soliton.
This can be seen on the level of dark soliton perturbation theory \cite{Kivshar1994},
resulting in an approximate particle-like equation of motion for the soliton center in the presence of a weak external potential.
This has been worked out for a dark soliton perturbed by a $\delta$-shaped impurity potential in \cite{Frantzeskakis2002},
with the result that the impurity is attractive (repulsive) to the soliton if it is attractive (repulsive) to the atoms in the condensate.
We have extended this analysis to our impurities of Gaussian shape (see appendix \ref{app} for the details) 
and find that the overall result persists: Dark solitons are effectively attracted by a Gaussian of $V_0 < 0$.
Specifically, the soliton center $x_0$ follows the equation of motion
\begin{align}
\dfrac{\mathrm{d}^2x_0}{\mathrm{d}t^2}=-\dfrac{\mathrm{d}W}{\mathrm{d}x_0},
\end{align}
where the effective potential $W(x_0)$ is predicted from perturbation theory, see appendix \ref{app} for the details.
Fig. \ref{effpot} shows the resulting effective potentials for varying widths of the Gaussian impurity (located at $x=0$), keeping the amplitude fixed. 
Clearly, a potential minimum for the soliton dynamics is found for all cases with $V_0 <0$, while it turns into an unstable maximum for $V_0 > 0$ (black circles in Fig.~\ref{effpot}).
In the vicinity of the fixed point $x=0$, the effective potential is strongly shaped by the Gaussian, while further away from the center it asymptotes to $\Omega^2 x^2/4$, half the bare potential of the trap, 
yielding the characteristic soliton oscillation at $\Omega/\sqrt{2}$ as expected in the absence of the Gaussian \cite{PhysRevLett.84.2298,DJF}.
\begin{figure}[!tb]
\centering
\includegraphics[width=0.45\textwidth]{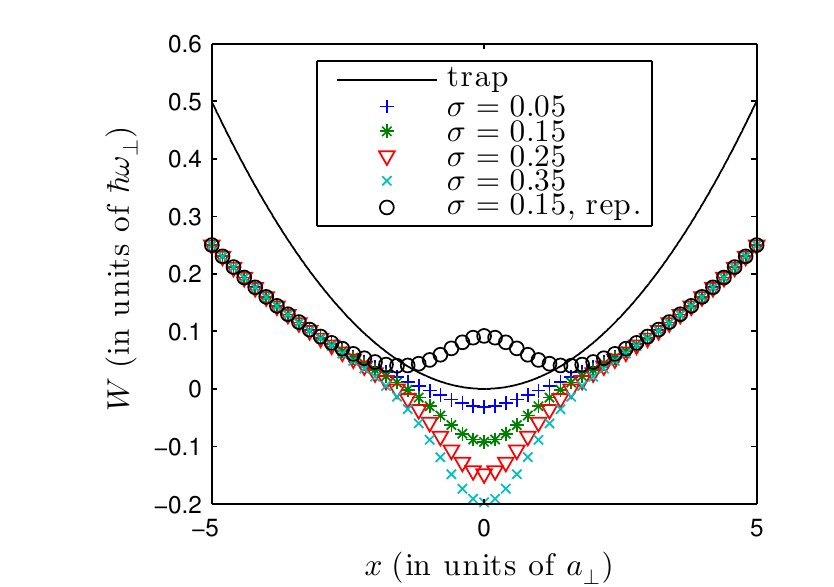}
\caption{{(Color online) Effective potentials $W(x_0)$ for the soliton position coordinate $x_0$ as obtained from soliton perturbation theory. 
In each case, the underlying atomic potential consists of a harmonic trap and a Gaussian of variable width $\sigma$ and fixed attractive amplitude $V_0=-1$ centered at $x=0$ (black circles for repulsive potential $V_0=1$), $\mu=1$ throughout.}}\label{effpot}
\end{figure}

This perturbative treatment suggests that for dragging along a dark soliton with a Gaussian impurity, one should choose the impurity as attractive for the atoms.
Consequently, we will focus on $V_0 < 0$ in the following. It should be noted that the parameters used in the rest of this work are mostly out of the range of validity of the soliton perturbation theory.

\section{Generating, pinning and dragging a dark soliton}\label{sec:generation}
\begin{figure}[!tb]
\centering
\includegraphics[width=0.45\textwidth]{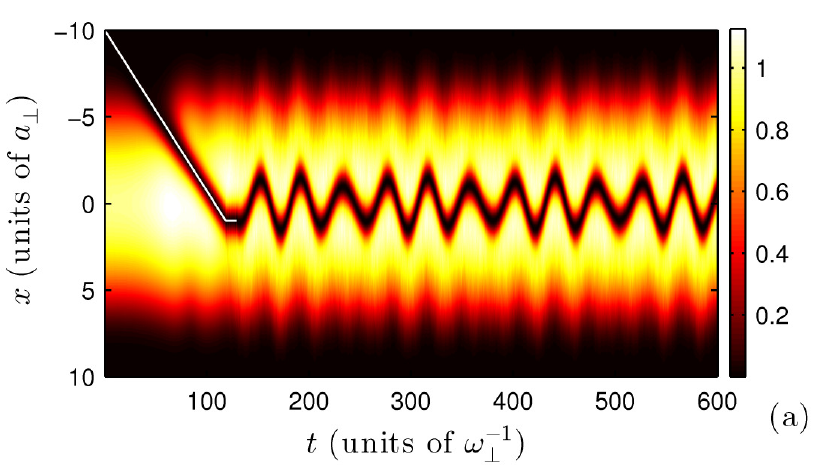}
\includegraphics[width=0.45\textwidth]{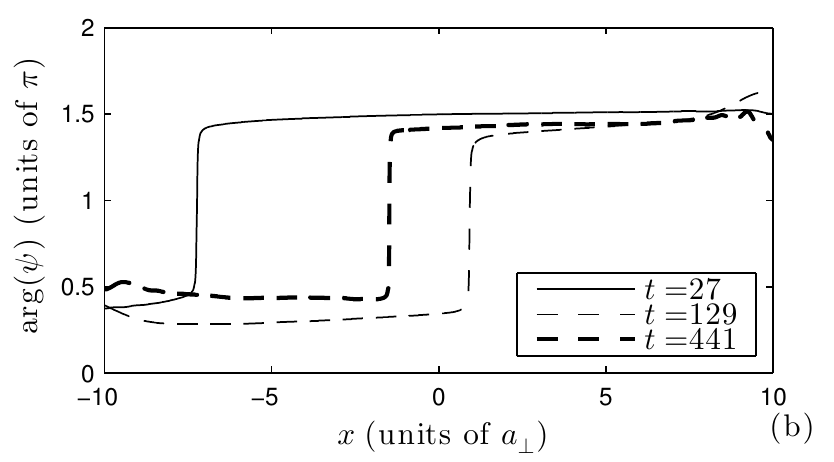}
\caption{{(Color online) Generating, dragging and releasing a dark soliton. The white line indicates the trajectory of the Gaussian impurity (parameters $V_0=-12, \sigma=0.1$). At $t=129$ the impurity is switched off. 
(a) Density $|\psi(x,t)|^2$. (b) Snapshots of the phase angle profile at different times.}}\label{gendrag}
\end{figure}

In this section, we demonstrate the possibility to generate, pin and drag along a dark soliton with an attractive Gaussian impurity entering the BEC cloud from outside.
An example of this is shown in Fig. \ref{gendrag}(a), displaying the spatio-temporal evolution of the atomic density under the influence of the impurity potential. 
The dimensionless chemical potential is chosen as $\mu=1$, corresponding for instance to a condensate of around 3300 sodium atoms under a transverse confinement of $\omega_\perp=2\pi\times 200$ Hz.
For these parameters, the resulting healing length in the center of the cloud is close to one micron.
The white line indicates the trajectory of the Gaussian that moves linearly into the BEC cloud towards $x=1$. After staying stationary at this point for a time interval of $\Delta t=10$, it is switched off. 
The generated soliton can be identified already at an early stage. When the impurity enters the cloud, the characteristic density minimum as well as the phase shift close to $\pi$ (see Fig. \ref{gendrag}(b)) are created almost immediately. 
The soliton follows the motion of the impurity and is dragged along towards $x=1$ where it is held for $\Delta t=10$. 
When the Gaussian potential is switched off, the soliton is released and starts to oscillate in the harmonic trap.
On a perfect Thomas-Fermi background, the frequency of this solitonic oscillation is expected to be $\Omega/\sqrt{2}$ \cite{PhysRevLett.84.2298,DJF}.
In the present simulation, the Gaussian impurity also slightly excites the collective dipole mode, causing a center-of-mass oscillation of the entire cloud at the trap frequency $\Omega$.
A two-sine fit to the soliton trajectory reveals a superposition between the particle-like soliton oscillation at $\Omega/\sqrt{2}$ and the collective dipole oscillation at $\Omega$,
validating the dark soliton character of the created excitation.

\begin{figure}[!tb]
\centering
\includegraphics[width=0.45\textwidth]{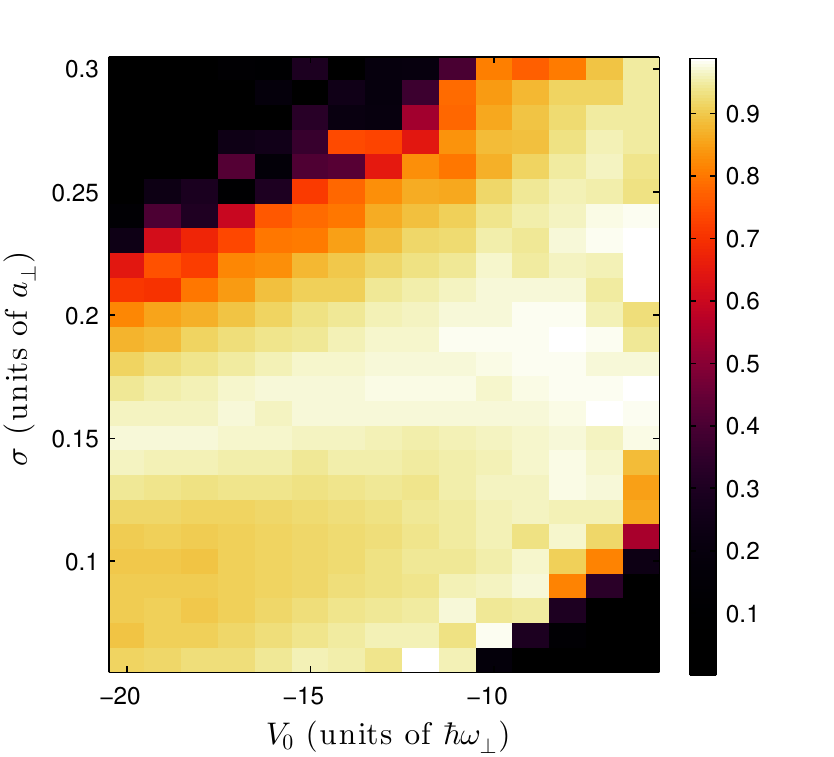}
\caption{{ Black soliton generation fidelity $\overline{S}$ as a function of the potential strength $V_0$ and width $\sigma$. The Gaussian enters the cloud at a velocity $v=0.0925$, the initial chemical potential $\mu =1$. }}\label{overlap}
\end{figure}

\begin{figure}[!tb]
\includegraphics[width=0.45\textwidth]{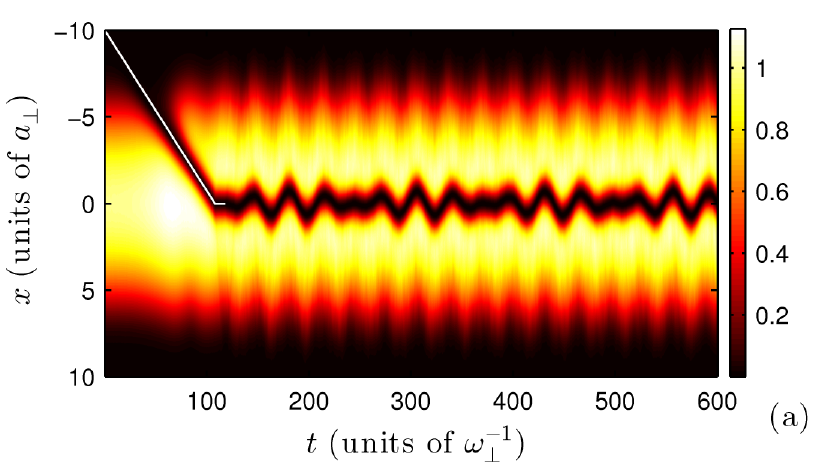}
\includegraphics[width=0.45\textwidth]{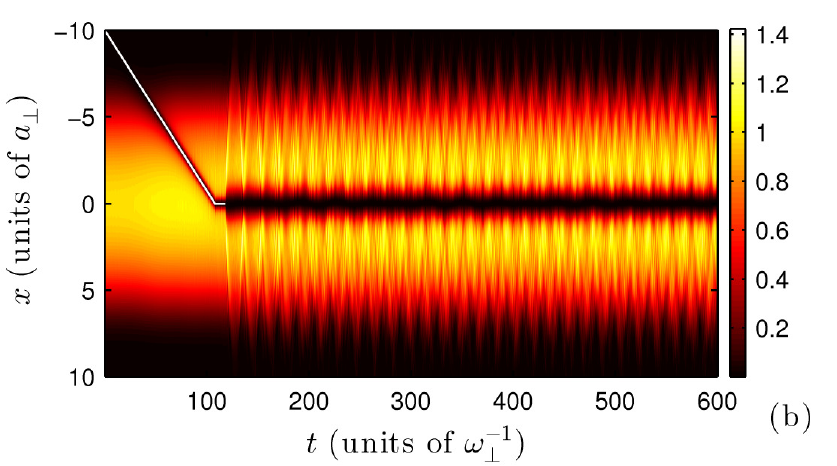}
\includegraphics[width=0.45\textwidth]{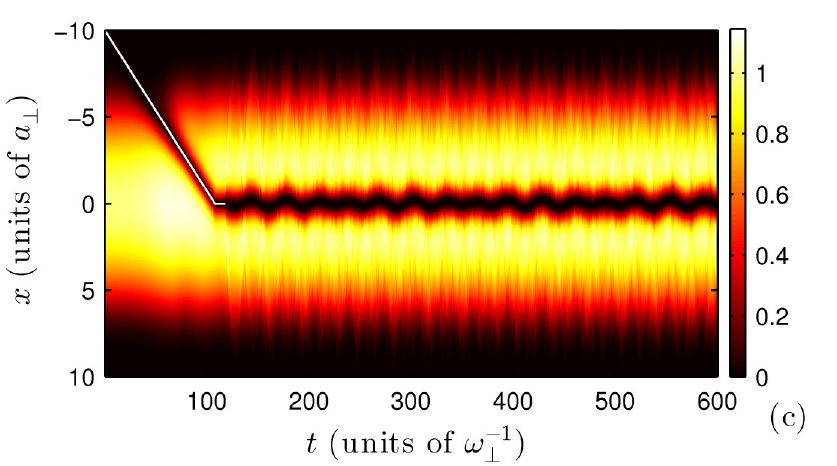}
\caption{{(Color online) Density evolution $|\psi(x,t)|^2$ for some of the simulations underlying Fig.~\ref{overlap}. The parameters of the Gaussians are $(V_0, \sigma)= (-12,0.1)$ (a), $(-7,0.28)$ (b) and $(-8,0.19)$ (c), respectively}.}\label{genex}
\end{figure}

We have confirmed in our simulations that the generation, immediate pinning and dragging of a dark soliton as shown in Fig.~\ref{gendrag} is successful in a wide range of parameters.
Specifically, while the velocity of the impurity is relevant (see also below), the exact trajectory is not, and the process works equally well with curved trajectories of the impurity and shorter or longer hold times prior to release (see, however, the discussion in Sec.~\ref{sec:instab}).
\\
To develop a fidelity measure for the dark soliton generation process, 
we need to identify a scenario in which a precise definition of the ideal desired outcome can be given. 
While on a homogeneous background there is a notion of a perfect grey (moving) soliton, 
an extension of this accounting for the inhomogeneous density induced by the trap is available on an approximate level only.
In contrast, the profile of a fully stationary (black) soliton in the trap can be obtained unambiguously by solving the stationary GPE.
Moreover, it can be expected that at parameters, at which an undisturbed initialization of a black soliton is possible, an off-center release from the impurity and
the subsequent acceleration due to the trap (similar as seen in Fig.~\ref{gendrag}) will produce a clean grey soliton.
Thus, we proceed to quantitatively evaluate the success of the soliton generation process by focusing on the preparation of a black soliton in the trap center.
To do so, we choose Gaussians of different parameters that enter the cloud at a given velocity and move towards $x=0$, are held there for a while and then turned off. 
For each of these runs, we compare the resulting final state to that of a stationary black soliton, our target state.
This target state $\phi_\text{BS}(x)$ (with the same squared norm $N$ as the wave function in the simulation) is computed separately by numerically solving the time-independent GPE.
Then, for any time $t$ after the potential has been turned off, we can calculate the overlap
\begin{align}
S(t)=\dfrac{1}{N^2} \bigg\vert \int \text{d}x \,\phi_\text{BS}^*(x) \psi(x,t)\bigg\vert^2.
\end{align}
This quantity is then averaged over a time-interval $[t_r,t_f]$, ranging from the release time $t_r$ (when the impurity is switched off) to a final time $t_f$:
\begin{align}
\overline{S}=\dfrac{1}{t_f-t_r}\int_{t_r}^{t_f}\mathrm{d}t\,S(t)\label{overlapeq}.
\end{align}
By the Cauchy-Schwarz inequality, $0\leq \overline S \leq 1$.
If the dynamics caused by the impurity results in a perfect stationary black soliton, one has $|\psi(x,t)|=|\phi_\text{BS}(x)|$ for all $t>t_r$ (taking advantage of the fact that the target state is stationary), and thus $\overline S = 1$ would correspond to a perfect fidelity of the creation process.
Smaller deviations from this indicate dynamics of the soliton and/or the bulk of the cloud after the impurity has been turned off, while $\overline S \ll 1$ suggests that the generation of a single dark soliton has completely failed.
Fig.~\ref{overlap} shows results for the fidelity $\overline S$ as a function of width $\sigma$ and height $V_0 <0$ of the Gaussian impurity.
The $1/e^2$ width of the Gaussian is given by $w=4 \sigma$ and ranges from $0.24$ to $1.2$ here. This is to be compared to the healing length in the center of the cloud given by $\xi \approx 0.7$.
The impurity moves towards the trap center on a linear trajectory, as in Fig.~\ref{gendrag}, at a velocity that is fixed to $v=0.0925$ in this set of simulations (measured in units of $a_\perp/\omega_\perp$). 
It is then suddenly stopped and held at $x=0$ for $\Delta t=10$, before being switched off.
The quantity $\overline S$ is then obtained by averaging $S(t)$ from the subsequent dynamics over an interval $t_f - t_r = 481$.
\\
Most notably, there is an extended parameter region of substantial fidelity $\overline S \gtrsim 0.9$.
A small or intermediate potential width $\sigma$ together with a large or intermediate potential strength $|V_0|$ is applicable for the controlled generation, dragging and holding of a single dark soliton. 
An exemplary plot at the same parameter values as in Fig.~\ref{gendrag} (but now with the final position of the Gaussian at $x=0$) is shown in Fig.~\ref{genex}(a) for the parameter set $(V_0=-12,\sigma=0.1)$. 
For this comparably narrow impurity, the final state is close to a stationary black soliton, but both the dipole mode and the soliton oscillation mode are slightly excited.
In contrast, turning to the parameters $(V_0=-7,\sigma=0.28)$, the fidelity is roughly the same, $\overline S \approx 0.94$, but the deviations from the black soliton state are of a different kind, see Fig.~\ref{genex}(b).
Here, the soliton itself is closer to stationary than in Fig.~\ref{genex}(a), but the background is excited at higher frequency modes and more disturbed by density waves.
The comparison of Figs.~\ref{genex}(a) and (b) illustrates that the fidelity $\overline S$ is sensitive to different types of remaining excitations around the target state (both particle-like oscillations of the soliton and collective oscillations in the bulk), and that one has some freedom in reducing either the particle-type soliton dynamics or the background excitations by tuning the parameters of the Gaussian.
In both regions of parameter space, the dark soliton characteristics of the induced density minimum are clearly observed; specifically, we have checked the oscillation at $\Omega/\sqrt{2}$ for off-centered release.
Finally, Fig.~\ref{genex}(c) shows a simulation at parameters $(V_0=-8,\sigma=0.19)$ that yields a particularly large fidelity of $ \overline S \approx 0.98$. Here, the evolution is similar to that shown in Fig.~\ref{genex}(b),
but the background excitations are further suppressed.

\begin{figure}[!tb]
\includegraphics[width=0.45\textwidth]{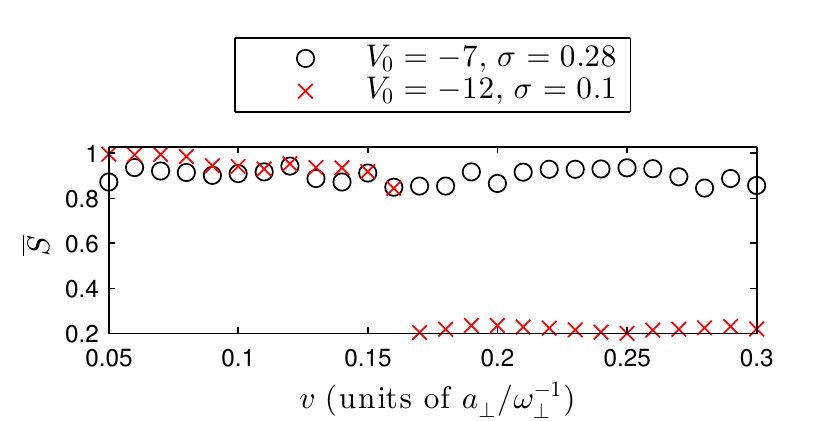}
\caption{(Color online) Black soliton creation fidelity as a function of the impurity velocity for two different sets of parameters of the Gaussian potential.}\label{velocity}
\end{figure}

So far, we have not addressed the role of the velocity at which the Gaussian defect is moved through the condensate.
For the two parameter sets $(V_0, \sigma)=(-12,0.1), (-7,0.28)$, we have performed simulations for a range of defect velocities $v$. 
The resulting fidelity $\overline S$ for varying $v$ is shown in Fig.~\ref{velocity}.
For both parameter sets, we find a robust plateau of large $\overline S$ at small $v$.
In particular for the deep and narrow Gaussian with $(V_0, \sigma)=(-12,0.1)$ the fidelity is close to $1$ for $v \lesssim 0.08$. 
Increasing the velocity, for this parameter set a sharp drop of $\overline S$ is observed at $v \approx 0.16$.
Beyond this velocity, in addition to a rather strong excitation of the background, further dark solitons are generated of which at most one is pinned by the Gaussian, which drastically reduces the overlap with the single black soliton state. 
This parameter regime could be of interest in its own right when moving the focus towards multiple soliton physics, but the high degree of control over the single soliton creation process is lost there.
In contrast, a comparable critical drop in the fidelity is not observed for the Gaussian with $(V_0, \sigma)=(-7,0.28)$, where $\overline S$ remains relatively large for an extended range of $v$.
For these parameters, going to even larger velocities than shown in Fig.~\ref{velocity} we observe a trend of decreasing $\overline S$, but caused by enhanced excitation of the background and the soliton oscillation mode, instead of multiple soliton formation.\\
In this work, we do not aim to explore the full velocity dependence and the dynamical details of the nucleation process in which the dark soliton is created when the defect enters the cloud.
Generally, the formation of excitations in a superfluid simultaneously exposed to a moving defect and a trapping potential (which leads to an inhomogeneous background density)
is an interesting and timely subject to study in its own right, see for instance the recent experiment on vortex shedding in quasi-2D condensates \cite{Kwon2015}.
Even on a homogeneous background, predicting the critical velocity above which a defect causes the creation of nonlinear excitations is an intricate problem (see for instance the extensive discussion of vortex nucleation in \cite{Barenghi2001}) that is subject to ongoing research \cite{Pinsker2014,Kunimi2015}. Some related results on soliton formation in 1D (mostly focusing on repulsive defects) are available, e.g. \cite{Hakim1997,Pavloff2002,Radouani2004,Kamchatnov2012}, 
but in our setup additional complications due to the inhomogeneous background density traversed by the defect, the attractive sign and the comparably soft boundary of the Gaussian potential \cite{Kwon2015} and possibly also the 1D reduction \cite{Fedichev2001} will require separate investigations that are beyond the scope of the present study.

\section{Instabilities for long hold times}\label{sec:instab}
\begin{figure}[!tb]
\centering
\includegraphics[width=0.45\textwidth]{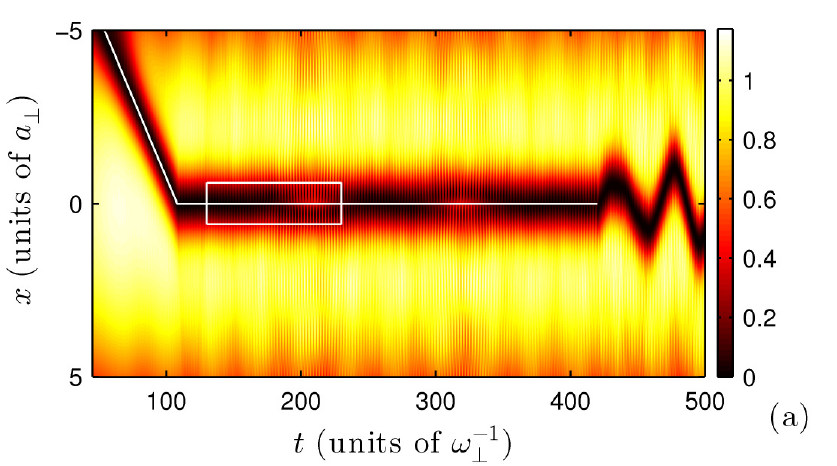}
\includegraphics[width=0.45\textwidth]{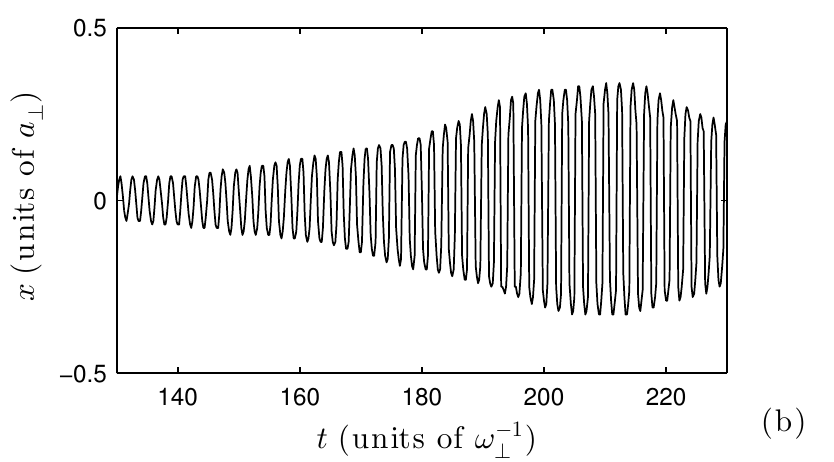}
\caption{{(Color online) Controlled creation of a dark soliton, followed by a long hold time. The parameters of the Gaussian are $V_0=-12$, $\sigma=0.08$, $v=0.0925$ (before reaching $x=0$). 
(a) Density evolution $|\psi(x,t)|^2$. The piece-wise linear trajectory of the impurity is marked by a white line. The white box highlights the first region of dynamical instability.
(b) Position of the density minimum (as a measure of the soliton center) as a function of time in the time interval marked by the box in (a).}}\label{instab}
\end{figure}

In the above simulations, the soliton was created and dragged by the moving impurity, then placed at a desired position in the trap and held there for a comparably short time ($\Delta t=10$) before being released. 
Substantially extending this hold time reveals an additional effect. For certain parameter values we observe instability phenomena in the dynamics of the pinned soliton, see Fig.~\ref{instab}. 
We find that the soliton performs micro-oscillations around the Gaussian potential during the hold interval (similar oscillations are observed in all our simulations).
Here, however, at $t \approx 140$ the amplitude of the micro-oscillations starts to increase strongly, before it decays again after $t \approx 210$. A similar increase and decrease is observed again at around $t\gtrsim 300$.
We conclude from this that long hold times may give rise to undesired effects when aiming for a stationary soliton.
Inadvertently releasing the soliton during a period of enhanced micro-oscillation will yield a comparably large momentum of its particle-like motion and correspondingly a large amplitude of its subsequent oscillations in the trap.
Thus, the hold interval may crucially affect the final state of the soliton preparation process.
Comparable dynamical instabilities of a dark soliton under the influence of a narrow external potential have been related to sound emission caused by the repeated asymmetric deformation of the oscillating soliton due to the impurity \cite{Parker2003a},
see also \cite{Parker2003,Yulin2003,Proukakis2004,Parker2004,Theocharis2005a,Allen2011}.

\begin{figure}[!tb]
\centering
\includegraphics[width=0.23\textwidth]{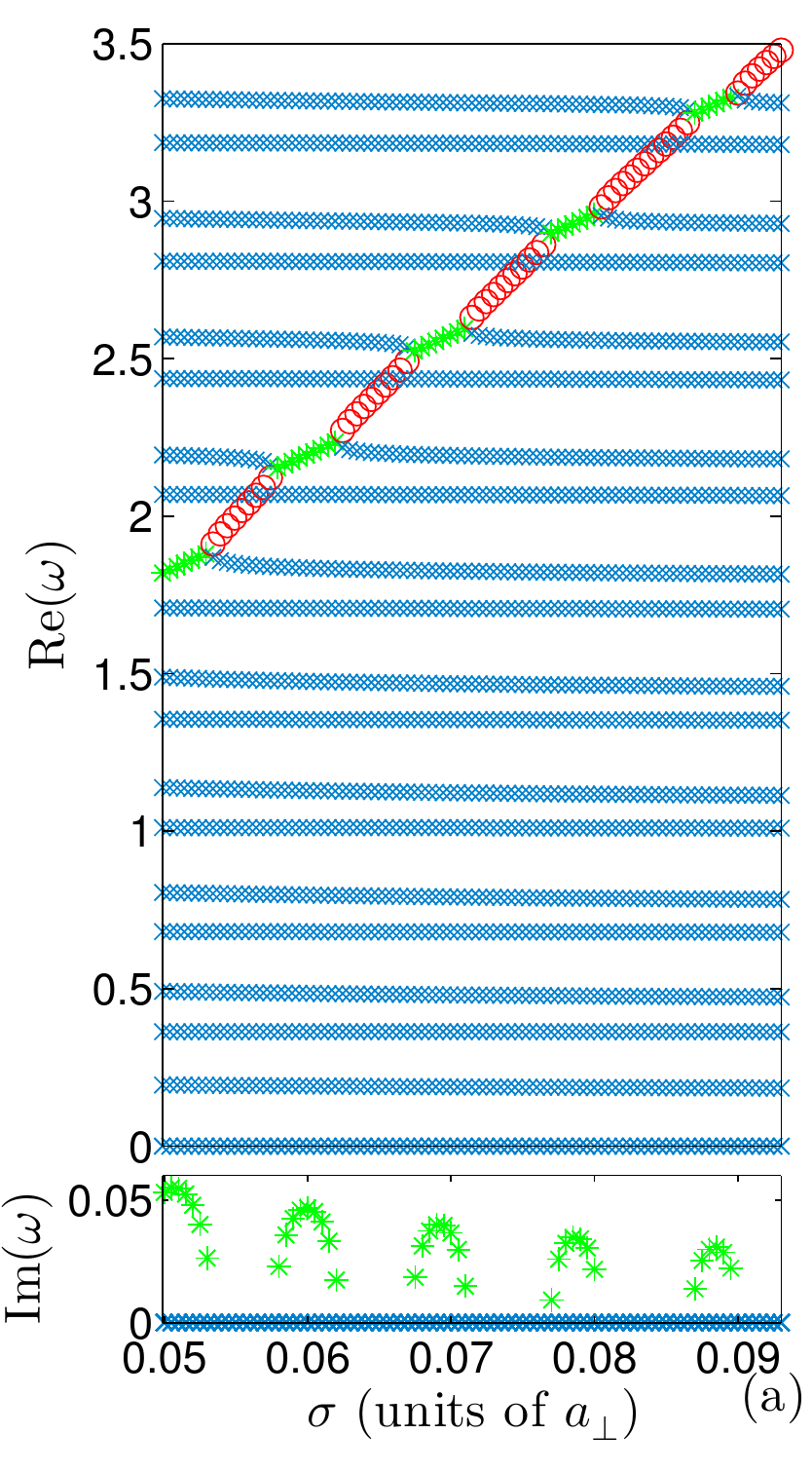}
\includegraphics[width=0.23\textwidth]{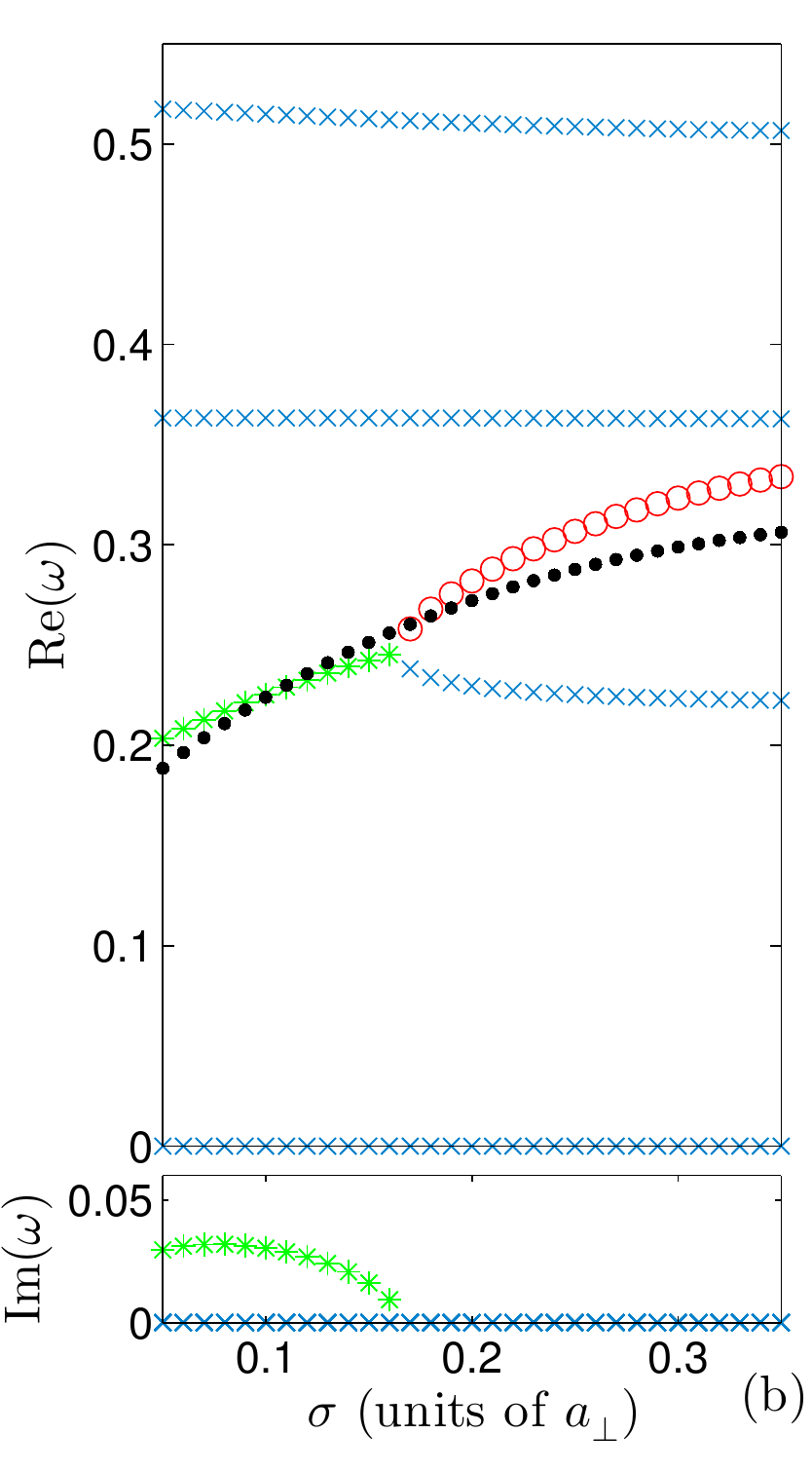}
\caption{{(Color online) BdG linearization spectra of the dark soliton state in the presence of a Gaussian potential in its center. All BdG frequencies are given in units of $\omega_\perp$.
Blue crosses, red circles and green asterisks denote the normal, anomalous and complex modes, respectively.
The chemical potential is fixed at $\mu=1.15$, yielding the same norm as the ground state at $\mu=1$. 
(a) $V_0=-12$, revealing a complex mode at $\sigma=0.08$, corresponding to the parameter set of Fig.~\ref{instab}.
(b) $V_0 = -0.25$, including also the linearization frequency predicted from the soliton perturbation theory (black dots).
}}\label{BdG}
\end{figure}

To obtain further insights into the dynamical instabilities due to the Gaussian potential, we employ a linearization analysis. 
Let us assume that the state that originates from the Gaussian entering the cloud and moving towards $x=0$ is close to the corresponding stationary black soliton state (now in the presence of the Gaussian). 
Then, information about its stability is encoded in the Bogoliubov-de Gennes (BdG) excitation spectrum, obtained by adding a small deviation
 $\delta \psi(x,t)=e^{-i \mu t} \left[ u(x) e^{-i \omega t}+ v^* (x)e^{i \omega^* t}\right]$
to the stationary dark soliton state, linearizing the GPE in $\delta \psi$ and solving the ensuing eigenvalue problem for $\omega$ \cite{book_kevrekidis}. Here, $\mu$ denotes the chemical potential of the stationary solution. 
Frequencies having nonzero imaginary part indicate an instability of the intial state as they induce exponential growth of a generic small perturbation. 
Such complex modes may emerge from collisions of normal and anomalous modes (modes with positive or negative energy/Krein signature, respectively) \cite{MacKay,book_kevrekidis}. 
Here, the BdG spectrum of the dark soliton state is expected to exhibit a single anomalous mode that is related to its particle-like motion \cite{DJF}.
If this becomes resonant with one of the background modes as a parameter is tuned (for instance the width of the Gaussian), this may lead to instability of the state.
Indeed, we observe these effects in the BdG spectrum of the black soliton state with a Gaussian potential placed in its center. We fix the norm of the wave function and the amplitude $V_0 = -12$ of the Gaussian 
to the same values as in Fig.~\ref{instab} and scan $\sigma$. The resulting spectrum as a function of $\sigma$ is shown in Fig.~\ref{BdG}(a) (by the Hamiltonian symmetry, if $\omega$ is in the BdG spectrum, then so are $-\omega$, $\omega^*$ and $-\omega^*$, so we can 
restrict to positive real and imaginary parts in the figure).
Here, even for moderately small $\sigma$ the large amplitude of the Gaussian strongly shifts the anomalous solitonic oscillation mode away from its value $\Omega/\sqrt{2}$ expected in the harmonic trap only. Increasing $\sigma$ leads to a further increase of the anomalous mode frequency,
causing subsequent collisions with background modes that result in complex quartets, signalling oscillatory instability. The width $\sigma=0.08$, as used in the simulation of Fig.~\ref{instab} indeed lies at the edge of such a region of instability.
Let us at this point return to the perturbative regime of small $|V_0|$ that was discussed in Sec. \ref{perturbth}.
Numerically calculating the BdG spectrum as a function of $\sigma$ also reveals the emergence of complex instability bubbles in this regime, see Fig.~\ref{BdG}(b) for an example.
The solitonic perturbation theory does not account for background excitation modes and is not capable of predicting these instabilities.
Interestingly, however, the linearization frequencies predicted from the perturbation theory (by linearizing the effective potential $W(x_0)$ around $x_0=0$) quite accurately capture the real part of the unstable BdG modes even in the instability regions.
In regions of stability, the frequency from the perturbative approach is close to the anomalous mode in the BdG spectrum, as expected. 
As is to be expected, the agreement slightly deteriorates for large $\sigma$, where the overall perturbation due to the Gaussian effectively becomes stronger.

\section{Perspectives and conclusions}\label{conclu}
\begin{figure}[!tb]
\centering
\includegraphics[width=0.45\textwidth]{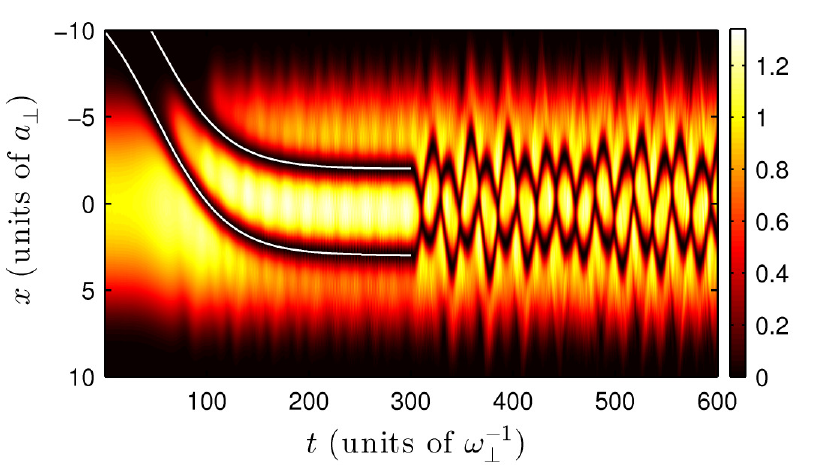}
\caption{{(Color online) Generating two solitons with two impurity potentials moving independently on the trajectories indicated by white lines. Colors encode the density $|\psi(x,t)|^2$. For both Gaussian potentials, $V_0=-14$, $\sigma=0.09$.}}\label{twosols}
\end{figure}

We theoretically investigated the possibility to controllably generate, drag, hold and release a dark soliton in a quasi-1D Bose-Einstein condensate using a Gaussian-shaped impurity as could be implemented with a focused laser beam. 
The time-dependent Gross-Pitaevskii equation containing the trap potential as well as the Gaussian impurity was propagated in time to obtain the spatio-temporal evolution of the condensate wave function when disturbed by the moving defect. 
On a perturbative level, we found that if the Gaussian is attractive (repulsive) to the atoms, then it is effectively attractive (repulsive) to the dark soliton as well,
thus suggesting the use of an attractive impurity (red-detuned focused laser) for holding and dragging dark solitons, in contrast to the pinning of vortices at repulsive barriers.\\
We demonstrated that by entering the atomic cloud with an attractive Gaussian one can create a single dark soliton that immediately sticks to the defect and can be controllably placed and released at a desired position in the condensate,
showing the expected characteristics of a dark soliton after release.
Detailed investigations revealed an extended range of model parameters (such as the width and amplitude of the Gaussian) for which this mechanism is successful, thereby underlining its robust nature.
As a drawback, the width of the Gaussian must be relatively small, comparable in size to the soliton healing length, requiring a much tighter focus than in previous experiments \cite{Engels2007}.
For instance, in the case of ${}^{23}\mathrm{Na}$ atoms and a transverse confinement of $\omega_\perp=2\pi\times 200$~Hz, a dimensionless value of $\sigma=0.2$ in our simulations 
(for which we observe particularly successful soliton generation and control) translates into a $1/e^2$ beam width of $w=4\sigma a_\perp \approx 1.2 \mu$m, close to the central healing length, while the wave function norm in our simulations translates into a relatively small number of around 3300 atoms. 
This is a challenging requirement, but not out of reach, given that optical systems with sub-micron resolution are already employed in present-day cold atom experiments \cite{Bakr2009,Zimmermann2011}.\\
Moreover, our studies suggest that long stationary hold times of the defect are not favorable for the controlled generation of black solitons, 
due to the possibility of dynamical instabilities that may lead to a spontaneous increase of the micro-oscillation amplitude of the soliton around the Gaussian.
This was related to corresponding complex modes arising in the Bogoliubov-de Gennes spectra, and linked to the linearization results from the perturbative approach.\\
The protocol for simultaneous generation and holding of the dark soliton as described here is appealing, since capturing an existing soliton in a BEC cloud would be a much more difficult task.
Even if a suitable pinning potential is available, catching the soliton requires information about its time-dependent position, which is hard to obtain given the destructive measurement schemes.
Furthermore, we point out that if more than one laser beam is available, the soliton creation scheme described herein can immediately be cascaded to generate two or more dark solitons at predefined positions as shown in Fig.~\ref{twosols}, cf. also \cite{Damski2001}.
The two dark solitons are created and trapped by their respective defect potential, and when released they exhibit the expected particle-like collision dynamics, cf. \cite{Weller2008,theocharis2010}.\\
In this work, we have not addressed the details of the soliton nucleation process occuring in the low density wings of the cloud and its dependence on the defect velocity.
This aspect of the problem promises to be an interesting topic for future studies.
Also, it would be desirable to perform simulations of the soliton creation and dragging protocol in the framework of the full three-dimensional Gross-Pitaevskii equation, 
checking for possible transverse excitation effects that are not captured within the dimensional reduction.

\begin{acknowledgments}
J. S. gratefully acknowledges support from the Studienstiftung des deutschen Volkes.
P. S. acknowledges financial support by the Deutsche Forschungsgemeinschaft via the contract Schm 885/26-1.
\end{acknowledgments}

\appendix
\section{Soliton perturbation theory}
\label{app}
In this appendix, we outline the derivation of the effective potential for the soliton in the harmonic trap, perturbed by a Gaussian impurity. 
We follow the presentation in \cite{Frantzeskakis2002}, but generalize the Dirac $\delta$-potential used in that work to a Gaussian as in Eq.~(\ref{potential}). 
By the same arguments as in \cite{Frantzeskakis2002}, the Thomas-Fermi-like background profile $u_b(x)$ of the condensate ground state is approximated by $u_b(x)=u_0+f_{\text{trap}}(x)+f_g(x)$, where $u_0$ is the maximum background amplitude ($u_0 = \sqrt{\mu}$ in the Thomas-Fermi limit considered here), $f_\text{trap}(x)$ accounts for the modified shape due to the harmonic trap and $f_g(x)$ incorporates the perturbation by the Gaussian impurity. Explicitly, $f_{\text{trap}}(x)=-\dfrac{1}{2u_0}V_{\text{trap}}(x)$ with $V_{\text{trap}}=\Omega^2 x^2/2$, as in \cite{Frantzeskakis2002}, and
\begin{align}
f_g(x)&= \notag \dfrac{V_0\sigma}{2}\sqrt{\dfrac{\pi}{2}} e^{2u_0^2\sigma^2}  \\ \notag
 &\times \left \{\left(-1+\text{erf}\left[\dfrac{\sigma}{\sqrt{2}}\left(2u_0+ \frac{x}{\sigma^2}\right)\right]\right)e^{2u_0x} \right. \\ 
& \,\, \left. +  \left(-1-\text{erf}\left[\dfrac{\sigma}{\sqrt{2}}\left(-2u_0+ \frac{x}{\sigma^2}\right)\right]\right)e^{-2u_0x} \right \}.\end{align}
Here, we can recover the result of \cite{Frantzeskakis2002} by taking the Dirac limit $V_0=b/(\sqrt{2\pi}\sigma)$ and $\sigma \rightarrow 0$ for a fixed $b$, resulting in $f_g(x)= -(b/2)\exp(-2u_0|x|)$.\\
With the generalized $u_b(x)$, we follow the further steps in \cite{Frantzeskakis2002}. The dynamics of the dark soliton on top of the Thomas-Fermi-like background is investigated with the Ansatz
\begin{align}
\psi(x,t)=u_b(x)e^{-iu_0^2t}v(x,t) \label{ansatz}
\end{align}
where $v(x,t)$ represents a dark soliton on this background. Inserting Eq.~(\ref{ansatz}) into the time-dependent GPE leads to a perturbed defocusing nonlinear Schrödinger equation for the soliton function $v$,
with a perturbation term that depends on $f_\text{trap}$ and $f_g$. 
Making an Ansatz for $v$ in the form of a dark soliton solution of the defocusing nonlinear Schrödinger equation, but with its position and phase angle slowly varying in time,
one can employ the adiabatic perturbation theory for dark solitons of \cite{Kivshar1994} to obtain the desired equation of motion of the soliton center $x_0 (t)$, which in our case reads as
\begin{align}\label{eq:effpot}
\dfrac{\mathrm{d^2}x_0}{\mathrm{d}t^2}&=\notag -\dfrac{1}{2}\dfrac{\mathrm{d}V_{\text{trap}}}{\mathrm{d}x}\bigg\vert_{x=x_0} \\
&+ u_0^3  e^{2u_0^2\sigma^2}\sqrt{\dfrac{\pi}{2}} \dfrac{V_0\sigma}{2}\int_{-\infty}^{\infty}\mathrm{d}x \left[ F_1(x) + F_2(x) \right] \notag \\
&=:- \dfrac{\mathrm{d}W}{\mathrm{d}x_0},
\end{align}
where the integrands
\begin{align*}
  F_1(x) &=\left(-1-\text{erf}\left[\dfrac{\sigma}{\sqrt{2}} \left(-2u_0+\frac{x}{\sigma^2}\right)\right] \right)e^{-2u_0x}  \notag \\
& \times \left\{\tanh [u_0(x-x_0)]-1\right\} \text{sech}^4 \left[u_0(x-x_0)\right], \\
  F_2(x) &=\left(-1+\text{erf}\left[\dfrac{\sigma}{\sqrt{2}} \left(2u_0+\frac{x}{\sigma^2}\right)\right] \right)e^{2u_0x}  \notag \\
& \times \left\{\tanh [u_0(x-x_0)]+1\right\} \text{sech}^4 \left[u_0(x-x_0)\right],
\end{align*}
From Eq.~\ref{eq:effpot}, we can numerically compute the effective potential $ W(x_0)$ (as shown in Fig.~\ref{effpot}) and the linearization frequency around its fixed point at $x_0 = 0$. 

 \bibliography{soliton}{}

\end{document}